\documentclass[aps,prl,twocolumn,superscriptaddress,showpacs]{revtex4-1}
\usepackage{graphicx}
\usepackage{bm}
\usepackage{color}
\usepackage{hyperref}
\usepackage[normalem]{ulem} %%to be deleted when removing \dawid

\begin{document}
\title{Peltier effect in strongly driven quantum wires}
%: limits of the linear response \\ or \\
%Nonequilibrium transport of entropy in quantum thermoelectric couple}
\author {M. Mierzejewski}
\affiliation{Institute of Physics, University of Silesia, 40-007 Katowice, Poland}
\author {D. Crivelli}
\affiliation{Institute of Physics, University of Silesia, 40-007 Katowice, Poland}
\author{P. Prelov\v{s}ek}
\affiliation{Faculty of Mathematics and Physics, University of Ljubljana, SI-1000 Ljubljana, Slovenia }
\affiliation{J. Stefan Institute, SI-1000 Ljubljana, Slovenia }

\begin{abstract}
We study a microscopic model of a thermocouple device with two connected correlated quantum 
wires driven by a constant electric field. In such isolated system we follow the time-- and 
position--dependence of the entropy density using the concept of the reduced density matrix. At weak driving, the initial changes of the entropy at the junctions can be described by the linear Peltier response. At longer times the quasiequilibrium situation is reached with well defined local temperatures which increase due to an overall Joule heating.  On the other hand, strong electric field induces nontrivial nonlinear thermoelectric response, e.g. the Bloch oscillations of the energy current. Moreover, we show for the doped Mott insulators that strong driving can reverse the Peltier effect. 
\end{abstract}
\pacs{71.27.+a,72.10.Bg,05.70.Ln}
%71.27.+a Strongly correlated electron systems; heavy fermions
%72.10.Bg General formulation of transport theory
%05.70.Ln 	Nonequilibrium and irreversible thermodynamics 
\maketitle

	Significant progress has recently been achieved in understanding the properties of strongly driven quantum many--body systems. The physics beyond the linear response (LR) regime  is interesting for basic research and important for future applications.  The underlying phenomena have become accessible to novel experimental techniques like ultrafast time--resolved spectroscopy of solids \cite{matsuda1994,dalconte12,rettig12,novelli12,gadermaier10,okamoto2010,cortes2011,kim12}
	 or measurements of relaxation processes in ultracold atoms driven far from equilibrium.  Most of theoretical studies  on transport beyond LR focus on charge currents driven by strong electromagnetic fields \cite{oka2003,jim2006,hasegawa2007,sugimoto2008,takahashi2008,
my1,my2,my3,lev2011,lev2011_1,eckstein2011,aron2012,amaricci11,einhellinger12} or heat/spin transport in electric insulators subject to a large temperature gradient \cite{Tomaz2011}. The thermoelectric phenomena beyond LR while important for power generation or cooling applications remain mainly unexplored, except for the specific case of non--interacting particles\cite{prosen2013}. First efforts in filling this gap have recently been reported in \cite{leijnse,kirchner} and \cite{sanchez2013} 
for quantum dots and mesoscopic systems, respectively. 
%In particular, the lowest order corrections to LR have been studied within the scattering theory in the latter paper. 

A thermoelectric couple (TEC) is the circuit build out of two different wires 
and is the basic device for heat--to--current 
conversion or  heat pumping. In this Letter we explore the behavior of a simple quantum model of 
an isolated TEC device connecting two wires of different materials with charge carriers being electrons and 
holes, respectively.  In a closed circuit the either weak or strong electric field can we introduce via induction.
We follow the real--time evolution of the TEC by solving the time--dependent Schr\"odinger equation.  
Since the system is isolated (decoupled from any thermal bath) the essential tool to investigate the 
local thermal properties is the concept of reduced density matrix (DM) of small subsystems. The latter allows to
study how the entropy density increases/decreases in different parts of the TEC. It allows also to specify 
the limits of the local equilibrium (LoE) regime. Although the Joule heating is the dominating non--linear effect 
it does not immediately break the LoE. On the contrary, the time-- and position--dependent temperature 
consistent with a canonical ensemble can be introduced also for moderate drivings far beyond the LR. 
We find that LoE persist up to much stronger fields, when the energy current starts to undergo the Bloch oscillations.
		
\begin{figure}
\includegraphics[width=0.45\textwidth]{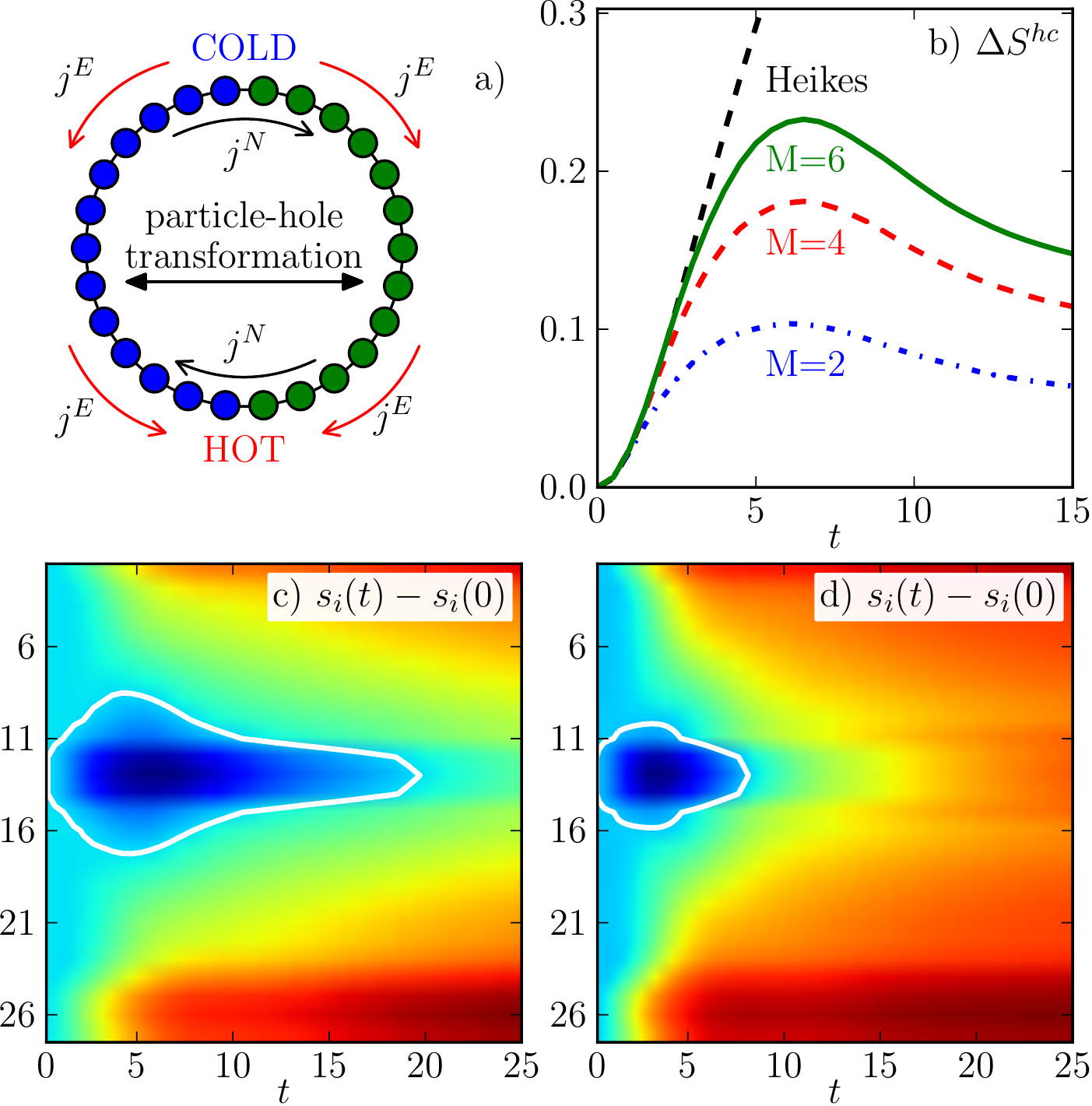}
\caption{(Color online) a): sketch of TEC. b): entropy difference between hot and cold junctions 
$\Delta S^{hc}$ for 
$\varepsilon_0=1.2$ calculated for $M$--site subsystems in comparison 
with Heikes Eq.~(\ref{ds}).
c) and d) show $s_i(t)-s_i(0)$ for $F=0.2$ and $F=0.4$, respectively,
for $M=4$, $\varepsilon_0=1.6$, with line denoting $s_i(t)=s_i(0)$. }
\label{fig1}
\end{figure}

We choose as the simple model for TEC the one-dimensional (1D) ring with $L$ sites and spinless but interacting 
fermions where different materials are modeled by site-dependent local potentials $\varepsilon_i$. 
Steadily increasing magnetic flux $\phi(t)$ induces an electric field $F=-\dot{\phi}(t)/L$, as 
described by the time--dependent Hamiltonian 
\begin{eqnarray}
H(t)&=&-t_0 \sum_i \left\{ {\mathrm e}^{i \phi(t)/L}\; c^{\dagger}_{i+1}c_i +{\mathrm h.c.} \right\} + \sum_i
\varepsilon_i n_i \nonumber \\
& &+ V \sum_i \tilde{n}_i \tilde{n}_{i+1} + W \sum_i \tilde{n}_i \tilde n_{i+2},
\label{ham}
\end{eqnarray}
where $n_i= c^{\dagger}_{i}c_i$ and $\tilde n_i=n_i -1/2$, $t_0$ is the hopping integral and periodic 
boundary conditions are used. $V$ and $W$ are repulsive interactions on nearest neighbors and 
next to nearest neighbors, respectively. The reason behind introducing $W$ is to stay away from 
the integrable case ($W=0$, $\varepsilon_i=\mathrm{const}$), which shows anomalous relaxation \cite{gge,Eckstein2012,Cassidy2011,my3} and charge transport \cite{u2,my1,my2,Tomaz2011,Marko2011,Sirker2009,Robin2011}. We model different wires
assuming a symmetric situation shown in Fig.~\ref{fig1}a, i.e. 
$ \varepsilon_i= -\varepsilon_0 $ and $\varepsilon_0$ for $i \in [1,L/2]$ and $i \in [L/2+1,L]$, respectively, while overall 
system is half-filled, i.e. the number of electrons $N_e=L/2$. Such a choice means that carriers in both wires are of
opposite character, i.e. they are electrons and holes, respectively. 

The dynamics of TEC is studied within a procedure described in Refs. \cite{my1,my2}. Initially $F=0$ and we generate a microcanonical state $|\Psi(0) \rangle$ for the target energy $E_0= \langle \Psi(0)| H(0) |\Psi(0) \rangle$ and small 
energy uncertainty $\delta^2 E_0=\langle \Psi(0)| [H(0)-E_0]^2|\Psi(0) \rangle $. Then the driving is switched 
on and the time evolution $ |\Psi (0) \rangle \rightarrow |\Psi (t) \rangle$ is calculated by the 
Lanczos propagation method \cite{lantime} applied to small time intervals $(t,t+\delta t)$.  
We use units in which $\hbar=k_B=t_0=1$. 

Since TEC is a composite object, its microscopic model may include several free parameters. 
With our choice of filling $N_e=L/2$ and $\varepsilon_i=\pm \varepsilon_0$ two parts of the TEC can be transformed to each other by a particle--hole transformation (see  Fig. \ref{fig1}). As a result, 
the concentration of fermions on one side of each junction is the same as the concentration of holes  
on other side $\langle n_{i} \rangle = 1 - \langle n_{L+1-i} \rangle$. It holds at any time provided that the same holds for $|\Psi(0)\rangle$.
In an isolated TEC the chemical potential $\mu$ is irrelevant, still within the grand canonical  
ensemble the considered $N_e=L/2$ case would correspond to $\mu=0$. 

The basic characteristics of the driven TEC are obtained from charge current $j^N_i =\langle J^N_i \rangle$ 
and energy current $j^E_i=\langle J^E_i \rangle$ defined from Eq.~(\ref{ham}) by the relations 
\cite{Naef97}
\begin{equation}
\nabla J^N_i \equiv J^N_{i+1}-J^N_{i}=i[n_i,H], \quad 
\nabla J^E_i = i[h_i,H],
\label{jne} 
\end{equation} 
where $H= \sum_i h_i $. So defined currents fulfill the continuity relations \cite{my1}
\begin{equation}
\frac{{\mathrm d} }{{\mathrm d} t} \langle n_i \rangle + \nabla j^N_i = 0 \label {c1}, \quad \quad
\frac{{\mathrm d} }{{\mathrm d} t} \langle h_i \rangle + \nabla j^E_i =  F(t) j^N_i \label {c2}.
\end{equation} 
Due to the imposed particle--hole symmetry in the chosen model, the charge currents are the same on 
both sides of the junctions 
$j^N_i=j^N_{L+1-i}$, while the energy currents flow in the opposite directions $j^E_i=- j^E_{L+1-i}$ 
(see Fig.~\ref{fig1}a). The latter property implies that magnitude of $\nabla j^E_i$ 
is particularly large at the junctions, what is the essence of the Peltier heating or cooling. 
The energy density changes also due to the Joule heating, as represented by the source term on 
the rhs. of Eq.~(\ref{c2}). However, the heating is of the order of at least $F^2$ while $\nabla j^E_i \propto F$. 

% dawid: added and removed a comma in 2 places
Since the initial state is a pure state with the corresponding DM 
$\rho(t=0)= |\Psi (0) \rangle \langle \Psi (0)|$ and the TEC is isolated from the surroundings, it stays 
in a pure state $|\Psi (t) \rangle \langle \Psi (t)|$ and the von Neumann entropy is identically zero. 
However, employing the concept of local reduced DM \cite{our2013} the entropy density can be obtained 
from DM of small subsystems of the TEC. For subsystems of $M$ consecutive lattice sites we calculate 
$\rho=\mathrm{Tr}_{L-M} |\Psi (t) \rangle \langle \Psi (t)|$ where the partial trace is taken over the 
remaining $L-M$ sites. Then, $S_i(t)=-\mathrm{Tr}_M(\rho \log \rho)$ is the local entropy and 
$s_i(t)=S_i(t)/M$ corresponding entropy density where $i$ 
labels the position of the subsystem within TEC. $s$ is thermodynamically relevant intensive quantity \cite{our2013,rigolast}
except for the low--energy regime where typically $s \propto M^{-1} $ according to the area laws \cite{Eisert2010}. Hence, we choose in this study the initial microcanonical 
states corresponding to high temperatures, i.e. initial $\beta(0)\simeq 0.3$. 
Furtheron we also set the size to largest available within our numerical approach, $L=26$.  

\begin{figure}
\includegraphics[width=0.45\textwidth]{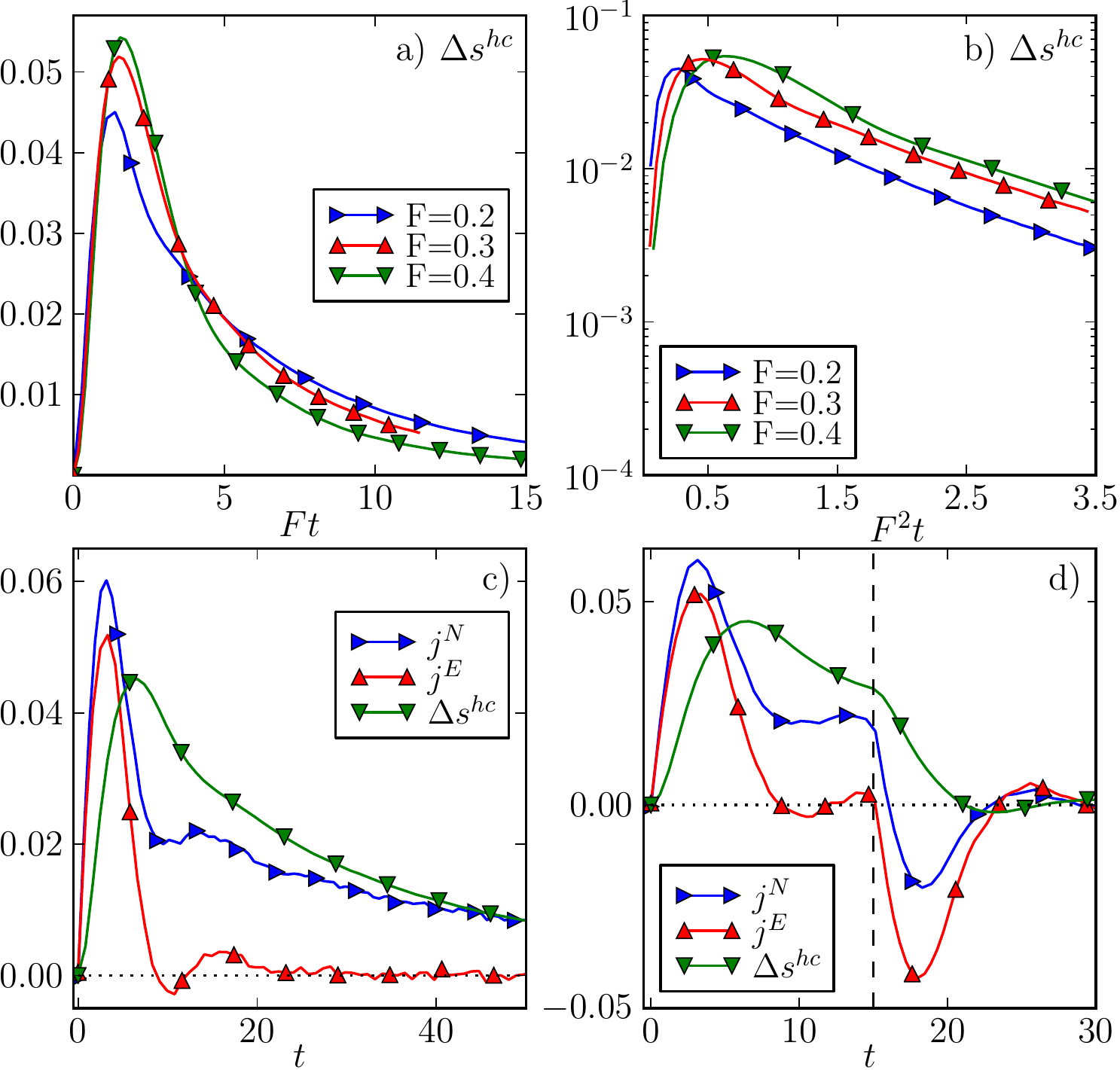}
\caption{(Color online) 
Results for $M=4$ and $\varepsilon_0=1.2$.
Difference of the entropy-densities  $\Delta s^{hc}$ is shown vs. a) $Ft$ and b) $F^2t$.
c) shows $j^N$ and $j^E$
in the middle of the left wire for $F=0.2$; d) the same but for $F$ 
switched off at $t=15$. }
\label{fig2}
\end{figure}

In order to identify the hallmarks of LoE we focus  on the weak--field regime. 
We consider metallic regime $V=1.4, W=1$ where the linear response functions are featureless \cite{u2}.
Figs. \ref{fig1}c and \ref{fig1}d show $s_i(t)$ for the TEC driven by $F=$const. Major changes of $s_i(t)$ are 
clearly visible at the junctions, i.e. at $i=$13 and 26. For short times $t<10$, $s_{13}(t)$ strongly decreases (we dub it 
the cold junction) while $s_{26}(t)$ strongly increases (hot junction). Due to particle--hole symmetry, driving does 
not affect the average concentration of fermions in subsystems
covering the junctions.
Therefore, the change of the entropy at the junctions must be due to genuine heating/cooling. Further support for this 
interpretation follows from Fig.~\ref{fig1}b, which shows the difference of the total entropies of subsystems 
which cover the hot and the cold junctions. Initially, the results are independent of $M$, indicating that
entropy is gained/lost mostly at the junctions consistently with Peltier heating $\dot{Q} = T \dot{S} = 
2 \Pi j^N$. At high $T$ we can employ the Heikes formula for each wire 
$ \Pi \simeq -\mu \sim \pm \varepsilon_0 $. The estimate is then
\begin{equation}
\Delta S^{hc} \equiv S^{\mathrm hot}(t)-S^{\mathrm cold}(t) \simeq 4 \beta(0) \int_0^t 
{\mathrm d t'} \varepsilon_0 j^N(t'), \label{ds}
\end{equation}
In the investigated regime the particle currents are determined by LR \cite{my1,my2}. 
Hence, the rate of the entropy gain/loss at the junctions is roughly proportional to 
$F$ as it is as shown in Fig.~\ref{fig2}a, well consistent with Eq.~(\ref{ds}).

Next we discuss the long--time regime shown in Fig.~\ref{fig2}b. Here
$\Delta s^{hc}(t)=\Delta S^{hc}(t)/M$ decays approximately as $\exp(-a F^2 t)$, 
where $a$ is independent of $F$. The same time--dependence has been found 
for particle current (see Fig.~\ref{fig2}c and Refs. \cite{my1,my2}) and explained as 
a result of the Joule heating. It has also been recognized as a hallmark 
of the quasiequilibrium (QE) evolution when $\rho$ is determined only by the 
instantaneous energy\cite{our2013}. Contrary to the case of homogeneous systems 
\cite{my1,my2,our2013}, the QE regime of TEC cannot be characterized 
by a single time--dependent $\beta(t)$.

An important property of the long time regime can be inferred from Fig.~\ref{fig2}c that shows 
$j^N$ and $j^E$ in the middle of the left part of TEC (far from the junctions). 
Initially, both currents show similar time--dependence, however $j^E$ vanishes for $t>10$ while $j^N$ remains large. In order to explain this result we recall that the in LoE regime both currents are driven by two independent forces: $F$ and $\nabla \beta$.
A particular combination of these forces may cause vanishing 
of $j^N$ (Seebeck effect) or $j^E$ (present case).
In order to explicitly show that vanishing of $j^E$  originates from compensation of two forces we instantaneously switch off one of them: the electric field. As shown in Fig.~\ref{fig2}d, the remaining force drives $j^E$ in the opposite direction. 
The magnitude of the resulting energy current is comparable with its
values during the initial evolution under $F \ne 0$. Below
we demonstrate that $\nabla \beta_i(t)$ 
is indeed the second driving force.

\begin{figure}
\includegraphics[width=0.45\textwidth]{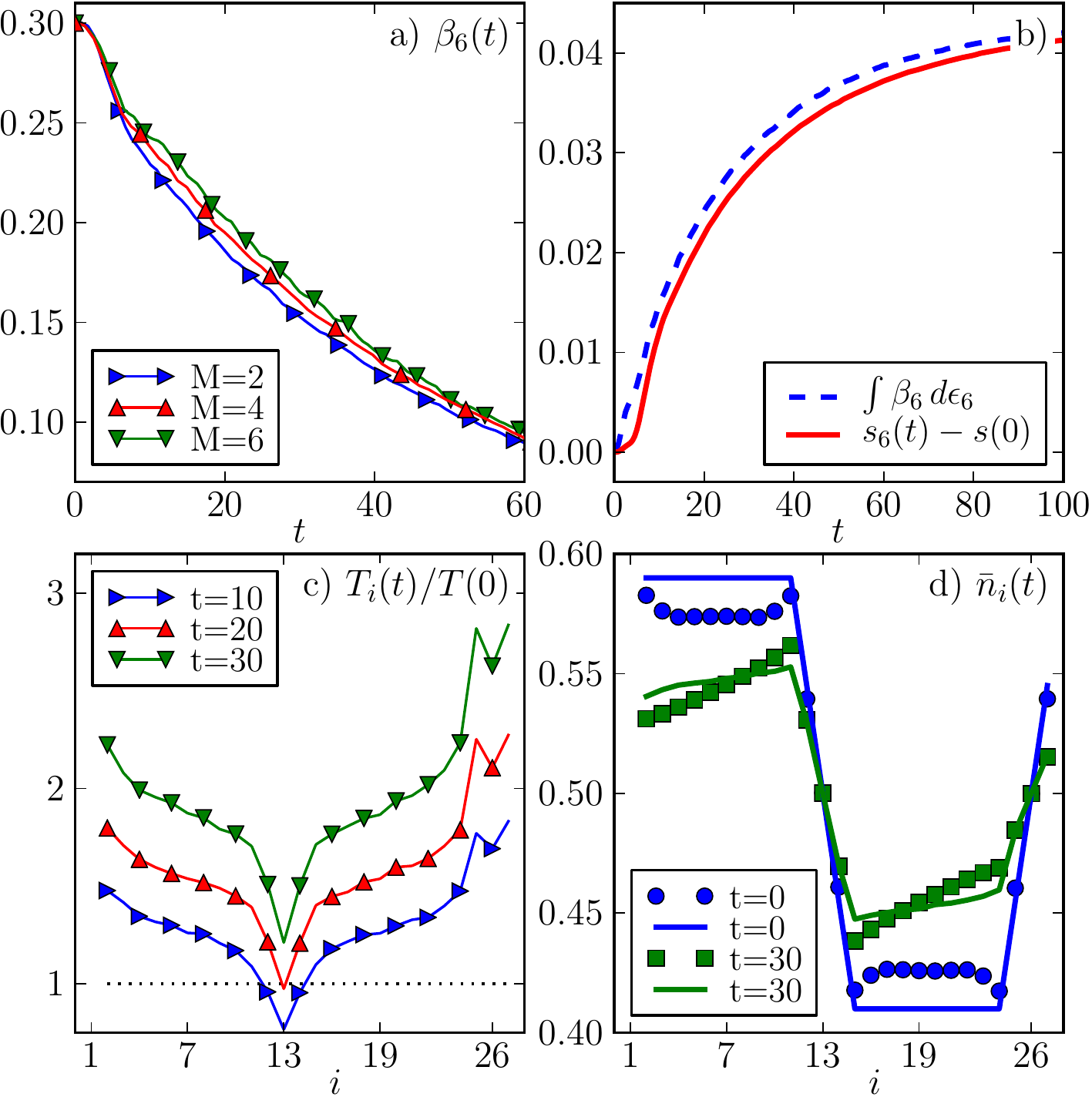}
\caption{(Color online) Results at $F=0.2$,
$\varepsilon_0=1.2$ for: a) $\beta_{i}(t)$ for $i=6$ (away from junctions), b) $s_{i}(t)-s_{i}(0)$ for $i=6,M=4$ determined
directly from $\rho$ and from $\int {\mathrm d} \epsilon_6 \beta_6 $, c) $\beta_i(t)$ for $t=10,20,30$, and 
d) $\langle n_i(t) \rangle $ for times $t=0$ and $t=30$ (points) compared with the
HTE result (lines).}
\label{fig3}
\end{figure}

It has been shown for a driven homogeneous wire that
$\rho$ is block--diagonal with respect to the number of particles
in the subsystem. In the QE regime
$\rho \propto \exp[ -\beta(t) H_{eff}]$ within each block \cite{our2013} and the spectrum 
$\{E_m \}$ of the effective Hamiltonian $H_{eff}$ is independent of $\beta$. 
Although for small subsystems $H_{eff}$ may significantly differ from $H$, one may still estimate
$\beta(t)$ without specifying explicit form of $H_{eff}$.
For the initial microcanonical state with known inverse
temperature $\beta(0)$ we determine the eigenvalues $\tilde{\lambda}_{m}$ of the largest block of $\rho$. 
Then a similar spectrum $\lambda_m$ is determined for a driven system in 
 a QE. Assuming the same $\{E_m\}$ one can then estimate 
 $\beta(t)/\beta(0) =\log(\lambda_m/\lambda_1)/ \log(\tilde{\lambda}_m/\tilde{\lambda}_{1})$.         
Fig.~\ref{fig3}a shows the resulting
$\beta_i(t)$ (averaged over $m \ne 1$) for the subsystem in the middle between
hot and cold junctions.  Being almost independent of $M$, $\beta$
is a well defined intensive quantity. 
%Similarly, $\beta$ can be introduced for all subsystems except at the junctions. 
Finally, we demonstrate that $\beta$ is consistent with the 2nd law of
thermodynamics. In Fig.~\ref{fig3}b we compare $s_i(t)-s_i(0)$
determined directly from $\rho$ with the integral
$\int_0^t {\mathrm d} \epsilon_i(t') \beta_i(t')$
where $\epsilon_i(t) =\langle h_i(t) \rangle$ is the energy density in the subsystem. Both
quantities are very close to each other. 
%A tiny shift between them develops at short times before the onset of QE. 
Therefore, we conclude that in
the QE regime one may introduce $\beta_i(t)$
consistent with the canonical ensemble as well as with equilibrium
thermodynamics. This consistency breaks down only for subsystem
covering one of the junctions. In fig. \ref{fig3}c	 we show snapshots of the temperature profiles $T_i$
for various $t$ in the QE regime. The temperature gradient is clearly visible, however there 
exists also an asymmetry between the change of $T_i$ at hot and cold junctions 
due to the heating effects.

In an inhomogeneous system $j^N \neq 0$ causes a
redistribution of particles within the TEC. This in turn may be  another (in addition to 
$F$) driving force for the transport of particles. In the investigated TEC this 
effect should be insignificant at least within the QE regime because of the 
particle--hole symmetry.
In order to confirm this expectation we plot in Fig.~\ref{fig3}d the spatial 
distribution of particles $\langle n_i (t) \rangle$ and compare it with the 
equilibrium high--temperature expansion (HTE), $n_i^{HTE}(t)= \bar{\varepsilon}_i \beta_i(t)/4 $, 
where $\varepsilon_i$ are  averaged  
over all sites of the subsystem. Indeed, the changes of $\langle n_i (t) \rangle$  can be 
reasonably explained as originating only from the time--dependence of $\beta_i(t)$. 

\begin{figure}
\includegraphics[width=0.45\textwidth]{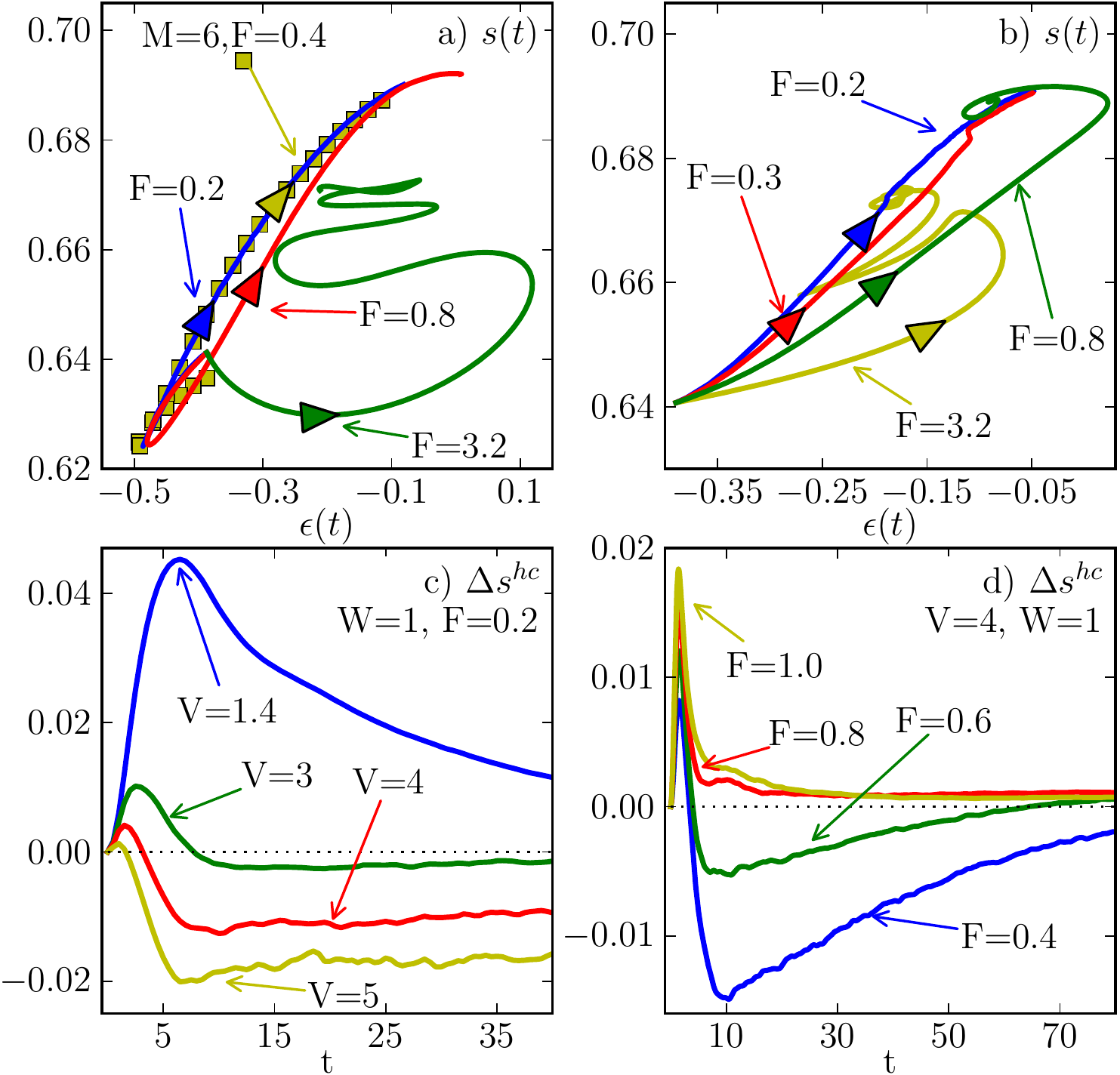}
\caption{(Color online) a) and b) show parametric plots $s_i(t)$ vs. $\epsilon_i(t)$ for cold and hot junctions, respectively.  
c) and d) show $\Delta s^{hc}(t)$ for various $V$ and $F$, respectively. Again, $M=4$ and $\varepsilon_0=1.2$.} 
\label{fig4}
\end{figure}

Next we concentrate on nonequilibrium phenomena
related with the operation of the TEC under strong $F$. 
The first one concerns the magnitude of $F$ which destroys the LoE. Since the TEC is 
spatially inhomogeneous LoE can be destroyed in certain parts of TEC while 
persisting in the other parts. 
% Hence, the question is not only when and but also where the LoE is destroyed. 
In the LoE regime, intensive quantities including $s_i(t)$ and $\epsilon_i(t)$, are uniquely determined by $\beta_i(t)$. 
%Since $s(\beta)$ and $\epsilon(\beta)$ are monotonic, there exists also one--to--one relation between $s_i(t)$ and $\epsilon_i(t)$.
%This relation holds in LoE independently of the initial conditions, $F$ or $M$. 
Such a universal relation is confirmed for $F \leq 0.4$ in Figs. \ref{fig4}a (cold junction) and \ref{fig4}b (hot junction). 
In the former case the curves for weak $F$ merge during
the entire evolutions, while in the latter case it happens only in the long--time regime after the nonequilibrium transient.
Results for $s_i(t)$ within the wires (not shown) are intermediate to the cases shown in Figs. \ref{fig4}a and \ref{fig4}b. 
Hence, one can observe that the LoE regime is broken first at the hot junction.
For large $F$, $\epsilon_i(t)$ starts to oscillate, while oscillations of $s_i(t)$ 
are rather limited. Therefore, the equilibrium relation between $\epsilon_i$ and $s_i$ is broken 
when the {\em energy current} $j^E_i(t)$ starts to undergo the Bloch oscillations. 
It is indicative to compare this result with recent finding for the driven homogeneous systems \cite{my1,eckstein2011} 
when the Bloch oscillations of the {\em particle current } $j^N_i(t)$ mark the onset of the nonequilibrium evolution.       
 
Finally we test the nonequilibrium response of TEC build out of two doped Mott insulators. 
Fig.~\ref{fig4}c shows the operation of TEC when the interaction $V$ is tuned 
from small (metallic) $V<2$ to large values $V \gg 2$ corresponding, close to half--filling,
to lightly doped Mott insulators. Such tuning reverses the dc flow of entropy (at longer $t$) and 
effectively interchanges the role of junctions (hot and cold, respectively).
%The junction dubbed previously (for metallic TEC) as hot becomes cold and vice versa. 
This effect is not unexpected being the result of changing the 
charge carriers close to half--filling from electrons in metallic regime to holes in the Mott-insulating regime. 
In contrast, results in Fig.~\ref{fig4}d are even more surprising. One can see that under 
strong driving $F > 0.5$, the Mott-insulating TEC operates in the same 
way as expected for generic metals, i.e. the current is again carried by electrons. 
Breaking of the Mott insulator {\em ground state} by strong $F$ has intensively been 
investigated during the last decade \cite{oka2003,oka2005,hasegawa2007,sugimoto2008,
takahashi2008,eckstein2010,zala2012,zala2012a} and explained mostly as a kind the Landau--Zener transitions 
from the dispersionless ground state to a dispersionful excited state. 
However in the present case, the breakdown concerns a {\em doped} 
Mott insulator and involves only excited states with rather high energy, so a proper explanation
remains a challenge.
 
In conclusion, we have studied a simple model of driven isolated TEC that can offer 
a useful and novel insight into several aspects of thermoelectric and nonequilibrium phenomena. 
Here, the concept of reduced (subsystem) DM is crucial for the discussion of increasing /decreasing entropy density,
local temperature and local equilibrium. Starting with an equilibrium state, we have shown that 
the onset of driving field $F$ first leads to local Peltier heating/cooling at junctions 
according to LR theory. Concerning the long--time regime of weakly/moderately driven TEC 
the behavior can be dubbed as "local quasiequilibrium". Similarly to the standard 
LoE one may introduce well defined $\beta_i(t)$. However, the changes of $\beta_i(t)$ originate 
not only from the energy and particle currents flowing within TEC, but also from the Joule heating 
due to external driving in analogy to QE in homogeneous systems \cite{my1} with homogeneous
$\beta(t)$. 

The presented method also allows to find the regions of evident departures
from LoE. In the metallic regime of the model, strong $F$ leads to the breakdown of the
relation between local temperature $T_i(t)$ and local energy $\epsilon_i(t)$ which
is incompatible with the notion of LoE. Even more dramatic are the effects
in the regime of doped Mott insulator 
where the charge carriers (within the equilibrium LR response) 
change the electron/hole character. Such systems are promising for the thermoelectric applications 
\cite{zlatic2013}. Here we find that large $F$ can even reverse the thermoelectric response.
 
\acknowledgements
Authors acknowledge stimulating discussions with Veljko Zlati\v{c}. 
This work has been carried out within 
the NCN project "Nonequilibrium dynamics of correlated
quantum systems". P.P. acknowledges the support by the Program P1-0044 and project J1-4244 of the Slovenian Research Agency.
 
\bibliography{bibliography.bib}

\end{document}